# Electronic and Thermodynamic Properties of the Amino- and Carboxamido-Functionalized C-60-Based Fullerenes: Towards Non-Volatile Carbon Dioxide Scavengers

Nadezhda A. Andreeva[1] and Vitaly V. Chaban[2]

(1) PRAMO, Saint Petersburg, Leningrad oblast, Russian Federation.

(2) P.E.S., Vasilievsky Island, Saint Petersburg, Leningrad oblast, Russian Federation.

**Abstract**. Development of new greenhouse gas scavengers is actively pursued nowadays. Volatility-caused solvent consumption and significant regeneration costs associated with the aqueous amine solutions motivate search for more technologically and economically advanced solutions. We hereby used hybrid density functional theory to characterize thermodynamics, structure, electronic and solvation properties of amino- and carboxamido-functionalized $C_{60}$ fullerene. $C_{60}$ is non-volatile and supports a large density of amino groups on its surface. Attachment of polar groups to fullerene $C_{60}$ adjusts its dipole moment and band gap quite substantially, ultimately resulting in systematically better hydration thermodynamics. Reaction of polyaminofullerenes with $CO_2$ is favored enthalpically, but prohibited entropically at standard conditions. Free energy of the $CO_2$ capture by polyaminofullerenes is non-sensitive to the number of amino groups per fullerene. This result fosters consideration of polyaminofullerenes for $CO_2$ fixation.

**TOC image**

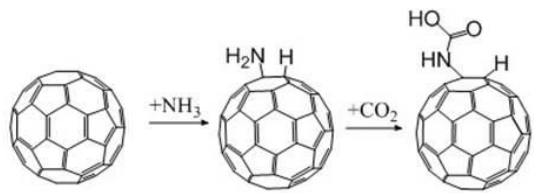

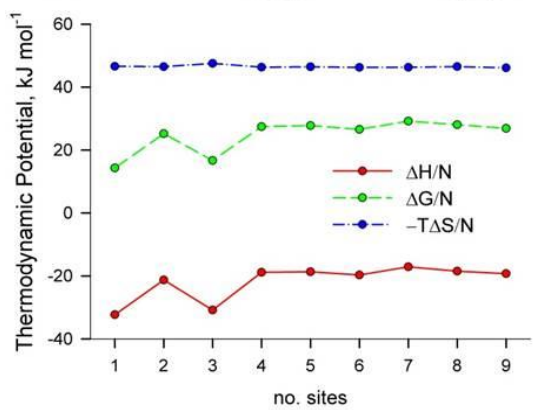

**Introduction**

Climate change constitutes one of the most significant global problems, which the humanity faces nowadays. According to some hypotheses, global warming is triggered by the concentration of the greenhouse gases (particularly carbon dioxide $CO_2$) increase in the atmosphere. As $CO_2$ has a number of industrial sources, many countries develop programs, which would allow them to decrease emission levels of $CO_2$ during the next decades. The major industrial methods to bind $CO_2$ are chemical and physical sorption, cryogenic distillation, membrane separation.[1-7] The most widely used of them is chemisorption by aqueous solutions of amines, such as monoethanolamine, diethanolamine, methyldiethanolamine as well as heterocyclic amines.[8,9] Reaction of amines and $CO_2$ yields carbamates, the reaction being reversible. The subsequent stage, called solvent regeneration, aims to shift the equilibrium leftwards by changing conditions (temperature, pressure) and regenerate reactants. Aqueous amines exhibit some well-documented drawbacks, such as high volatility and relatively high regeneration costs.[10] High volatility of amines in their aqueous solutions at room temperature is a serious problem, since it results in a significant consumption of the working substance during every cycle of its usage. New non-volatile amino derivatives would solve this exploitation problem. It is also important to maintain competitive capacity and lower regeneration costs. Increasing the density of amino groups per mass unit of the $CO_2$ scavenger should allow ever higher $CO_2$ capacities.

Development of new approaches and working substances to capture $CO_2$ is currently underway.[11-13] For instance, Zhuang and co-workers[14] reported amino-functionalization of the polypropylene fiber and encouraging $CO_2$ capture performances of the resulting products. Einloft and co-workers[15] linked a variety of functional groups to cellulose and observed their affinity to $CO_2$. Cellulose is a cheap product, whose cost makes it a natural starting point to synthesize versatile low-cost, low-volatility $CO_2$ scavengers. Decoration of the inexpensive polymers with amino groups is also pursued actively.[16-18]

Fullerenes are an allotropic form of carbon exhibiting a peculiar electron acceptor behavior, possessing unique physical properties, and offering unprecedented applications.[19-24] Fullerenes constitute a large groups of all-carbon compounds, $C_{20}$ to $C_{720}$, which are possible due to the chemical property of carbon to maintain fused pentagons and hexagons. Thanks to their curved surface, fullerenes are polarizable and reactive giving rise to rich chemistry.[25] Covalent functionalization opens numerous avenues to combine physical properties of fullerenes with those of the grafted molecules. Despite their electronic polarizabilities and reactivities, most fullerenes exhibit outstanding kinetic stabilities. For instance, $C_{60}$ (also widely known as buckminsterfullerene and buckyball, Figure 1) sublimates without decomposition at 873 K.[26] Pristine fullerenes are insoluble in water, which is generally a problem for many prospective applications. Due to their large molecular masses, fullerenes are non-volatile, unlike amines and many other organic substances.

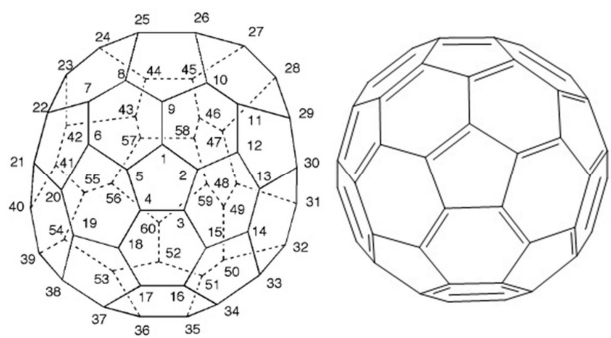

Figure 1. (Left) Atom numbering in ($C_{60}$-$I_h$)[5,6]fullerene following the IUPAC recommendations. (Right) Spatial model of $C_{60}$.

Recent chemical modeling showed that calcium-doped $C_{60}$ attracts greenhouse gases ($CO_2$ and $N_2O$) much more efficiently,[27] as compared to the non-functionalized buckyball. Up to five $CO_2$ molecules get attached to each calcium atom driven by an electrostatic attraction. The dipole moment of $C_{60}$ increases greatly upon adsorption of calcium. Adsorption of $N_2O$ is even stronger, since this molecule possesses a non-zero dipole moment. Lin and co-workers amino-functionalized fullerene and carbon nanotube.[28] According to these authors, the sites forming a larger pyramidalization angle exhibit higher reactivities. The most thermodynamically favorable sites for the amino group attachment are located at the pentagon-pentagon fusions of $C_{50}$. In the case of carbon nanotube, the 7-5-5-7 structure defect on the sidewall and the junction of two tubes results in the most stable amino-functionalized products. Density functional theory reveals that the reaction energy of the addition of methylamine onto $C_{60}$ or closed caps of CNTs is rather low. The reaction at a few sites appears even exothermic. Understanding of thermochemistry is necessary to foster functionalization experiments.

Since $C_{60}$ is non-volatile and supports versatile chemical reactions on its surface, it is interesting to consider it as a robust substrate for the $CO_2$ scavengers, such as amines. $C_{60}$ can be amino-functionalized, as shown in the experiments and computations previously.[29-32] Subsequently, $C_{60}H_n(NH_2)_n$ can capture $CO_2$ through chemisorption. Thanks to a significant surface area, $n$ in $C_{60}H_n(NH_2)_n$ is expected to be sufficiently large to compete with other $CO_2$ scavengers per unit of mass or volume. It must be stated, however, that the present production costs of fullerenes remain too high to compete with aqueous amine solutions and other technologies. In the present work, we investigate electronic, structure, and standard thermodynamic properties of a series of $C_{60}H_n(NH_2)_n$ and $C_{60}H_n(NH_2)_n(CO_2)_n$ to characterize feasibility of the $CO_2$ chemisorption reaction, as depicted in Figure 2.

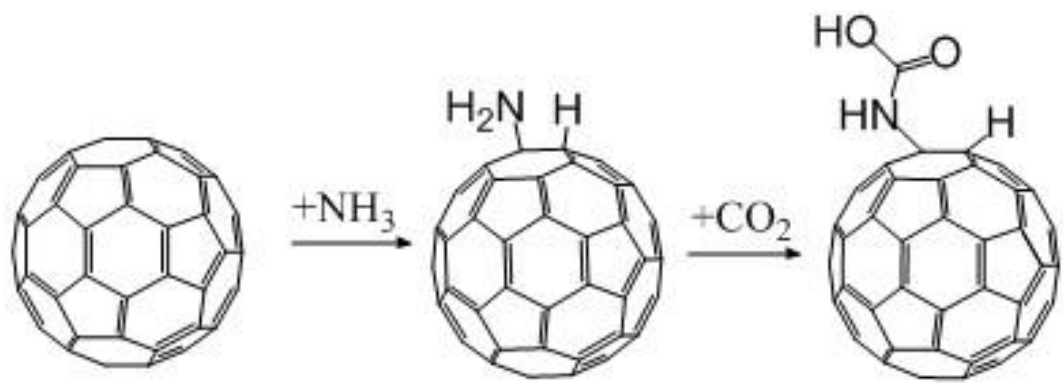

Figure 2. Schemes of amino- and carboxamido-functionalization starting from $(C_{60}\text{-}I_h)[5,6]$fullerene.

**Methodology**

We considered $C_{60}H_n(NH_2)_n$ and $C_{60}H_n(NH_2)_n(CO_2)_n$ with *n* = 1, 2, 3, 5, 6, 8, 9 (Table 1). No more than one amino group was grafted to any single hexagon or pentagon on the surface of $C_{60}$. Thermodynamic potentials, enthalpy $\Delta H^0$, entropy $\Delta S^0$, and Gibbs free energy $\Delta G^0$ at standard temperature and pressure for the studied reactions (Figure 2) were computed by density functional theory (through molecular partition functions) using the gas-phase approximation.[33] The harmonic oscillator behavior was assumed. The Hess rule was used to compute standard thermodynamic potentials for the studied chemical reactions by subtracting the corresponding potential of all reactants from the corresponding potential of all products.

Table 1. Formulas and names of the amino- and carboxamido-functionalized $C_{60}$-based fullerenes.

| no. | Brutto Formula | amino/carboxamido positions* | hydride positions* |
|---|---|---|---|
| 1 | $C_{60}$ | none | none |
| 2 | $C_{60}H(NH_2)$ | 1 | 9 |
| 3 | $C_{60}H_2(NH_2)_2$ | 1,10 | 9,11 |
| 4 | $C_{60}H_3(NH_2)_3$ | 1,6,19 | 5,9,20 |
| 5 | $C_{60}H_4(NH_2)_4$ | 1,10,29,47 | 9,11,28,48 |
| 6 | $C_{60}H_5(NH_2)_5$ | 1,6,19,39,53 | 5,9,20,38,54 |
| 7 | $C_{60}H_6(NH_2)_6$ | 1,6,10,19,39,53 | 5,9,11,20,38,54 |
| 8 | $C_{60}H_7(NH_2)_7$ | 1,6,10,19,29,39,53 | 5,9,11,20,28,38,54 |
| 9 | $C_{60}H_8(NH_2)_8$ | 1,6,10,19,29,39,47,53 | 5,9,11,20,28,38,48,54 |
| 10 | $C_{60}H_9(NH_2)_9$ | 1,6,10,19,29,39,47,49,53 | 5,9,11,20,28,38,48,54,59 |
| 11 | $C_{60}H(NH_2)(CO_2)$ | 1 | 9 |
| 12 | $C_{60}H_2(NH_2)_2(CO_2)_2$ | 1,10 | 9,11 |
| 13 | $C_{60}H_3(NH_2)_3(CO_2)_3$ | 1,6,19 | 5,9,20 |
| 14 | $C_{60}H_4(NH_2)_4(CO_2)_4$ | 1,10,29,47 | 9,11,28,48 |
| 15 | $C_{60}H_5(NH_2)_5(CO_2)_5$ | 1,6,19,39,53 | 5,9,20,38,54 |
| 16 | $C_{60}H_6(NH_2)_6(CO_2)_6$ | 1,6,10,19,39,53 | 5,9,11,20,38,54 |
| 17 | $C_{60}H_7(NH_2)_7(CO_2)_7$ | 1,6,10,19,29,39,53 | 5,9,11,20,28,38,54 |
| 18 | $C_{60}H_8(NH_2)_8(CO_2)_8$ | 1,6,10,19,29,39,47,53 | 5,9,11,20,28,38,48,54 |
| 19 | $C_{60}H_9(NH_2)_9(CO_2)_9$ | 1,6,10,19,29,39,47,49,53 | 5,9,11,20,28,38,48,54,59 |

* For reference, the IUPAC name of $C_{60}H_8(NH_2)_8$ is 1,6,10,19,29,39,47,53-octaamino-5,9,11,20,28,38,48,54- octahydride -($C_{60}$-$I_h$)[5,6]fullerene. The IUPAC name of $C_{60}H_5(NH_2)_5$ is 1,6,19,39,53-pentacarboxamido-5,9,20,38,54- pentahydride -($C_{60}$-$I_h$)[5,6]fullerene.

To compare an effect of mutual location of the grafted groups, two isomers of composition $C_{60}H_n(NH_2)_2(CO_2)_2$ were considered. One isomer contains amino groups, and consequently, carboxamide groups, within the neighboring carbon hexagons. Another isomer contains amino groups within the opposite carbon hexagons, so that to remove or minimize dipole moment of the resulting chemical structure.

Because the investigated molecules are of significant size, application of a high-level post-Hartree-Fock electron-correlation methods or composite thermochemistry methods is impossible, both due to the memory requirements and the processor speed requirements. In this situation, the M11 hybrid density functional theory (HDFT)[34] was used to obtain optimized wave functions. According to the developers, M11 provides better properties, as compared to the previously devised functionals of this line. Upon applying DFT, in lieu of the electron-correlation methods, one should expect somewhat lower accuracy than that for small molecules, but realize that the present choice is the only available way to obtain thermodynamics of the investigated reactions. The atom-centered split-valence double-zeta polarized basis set 6-31G(d) was used. All electrons were considered explicitly. The wave function energy convergence criterion at every self-consistent-field iteration was set to $10^{-8}$ Hartree.

Population analysis was performed using the conventional Mulliken scheme. The Mulliken partial charges are directly related to the electron population levels. Solvation free energies in water and n-octanol were assessed the solute charge density and continuum solvent model, as exemplified previously by Marenich and Truhlar.[35,36] Gaussian09 was used to conduct electronic-structure simulations.[37] Gabedit[38] and VMD[39] packages were used to manipulate molecules.

**Results and Discussion**

Attachment of the amino and carboxamide groups to $C_{60}$ is similar to their attachment to alcohols and alkanes. The carbon-nitrogen ($NH_2$) covalent distance in both ethanolamine (MEA) and polyamino-$C_{60}$ is 1.47 Å. After the carboxyl group is linked, this distance decreases to 1.45 Å in MEA, but to 1.46 Å in $C_{60}$-based polycarbamates. The nitrogen(amine)-carbon(carboxyl) distance is 1.36 Å in $C_{60}$-polycarbamates and 1.38 Å in ethanol carbamic acid. The $C(C_{60})$-$C(C_{60})$-N(amine) angles range 106-112 degrees when the amino group is connected to the five-membered ring of $C_{60}$ and 105-110 degrees when the six-membered ring of $C_{60}$ is functionalized. This same angle is MEA is 110 degrees, but increases to 114 degrees in ethanol carbamic acid. The $C(C_{60})$-N(amine)-C(carboxyl) angles are 124-128 degrees in all $C_{60}$-polycarbamates. This same angle in ethanol carbamic acid is 126 degrees. Thus, all detected structural differences between aminated $C_{60}$ and MEA are insignificant. The amino-functionalization of $C_{60}$ does not result in any additional strains of the $C_{60}$ fullerene structure and does not undermine stability of the carbonaceous backbone.

The standard thermodynamic potentials for amino- and carboxamido-functionalization vs. the number of attached groups are summarized in Figure 3. The standard Gibbs free energies are slightly positive that is expected at 1 bar and room temperature. Elevation of pressure would shift an equilibrium towards non-gaseous products. In all cases, an impact of the additional groups is relatively small, whereas certain free energy decrease at higher amino group loadings is observed. The observed trend is in concordance with the previous report by other researchers.[28] The major physical reason behind this trend lies within favorable dipole-dipole interactions of the grafted groups. Free energy decrease is not uniform and mutual location of the groups also matters. For instance, formation of $C_{60}H_5(NH_2)_5$ is more favorable than formation of $C_{60}H_4(NH_2)_4$ and $C_{60}H_6(NH_2)_6$. On the contrary, formation of $C_{60}H_5(NH_2)_5(CO_2)_5$ is somewhat less favorable than formation of $C_{60}H_4(NH_2)_4(CO_2)_4$ and $C_{60}H_6(NH_2)_6(CO_2)_6$. Strong electrostatic interactions between and the -NH group and the -COOH group play an important role in stabilizing the functionalized derivatives. This effect is more pronounced in the case of higher degrees of functionalization. Further investigation of the electronic properties of $C_{60}H_n(NH_2)_n$ and $C_{60}H_n(NH_2)_n(CO_2)_n$ was used to corroborate the thermodynamic observations. In the case of amino-functionalization, entropy plays a modest role, as compared to enthalpy, which also does not essentially depend on the number of the grafted groups. In turn, the fraction of entropy in the $CO_2$ chemisorption is significant, confirming the conclusion of the previously reported calculations.[40] Since the chemisorption reaction decreases the total number of molecules in the simulated system, the corresponding entropy alteration is unfavorable, TdS < 0. Pressure increase shifts the equilibrium of the carbamate formation reaction rightwards, since the total number of molecules in the system decreases.[40,41] Relatively high free energies of the carbamate formation imply accordingly low regeneration energy costs, i.e. those for shifting the reaction equilibrium leftwards.

A relatively small effect of the number of grafted groups on the reaction Gibbs free energy motivates to consider higher -$NH_2$ contents in $C_{60}H_n(NH_2)_n$. Thermodynamic calculations revealed that $C_{60}$ constitutes an interesting substrate to host groups with a high affinity to $CO_2$. Fullerenes are non-volatile and thus, unlike in aqueous amines, the $CO_2$ scavenger losses associated with their potentially numerous sorption/desorption cycles are minimal. Furthermore, $C_{60}H_n(NH_2)_n$ and $C_{60}H_n(NH_2)_n(CO_2)_n$ are sufficiently stable at standard conditions (Figure 3).

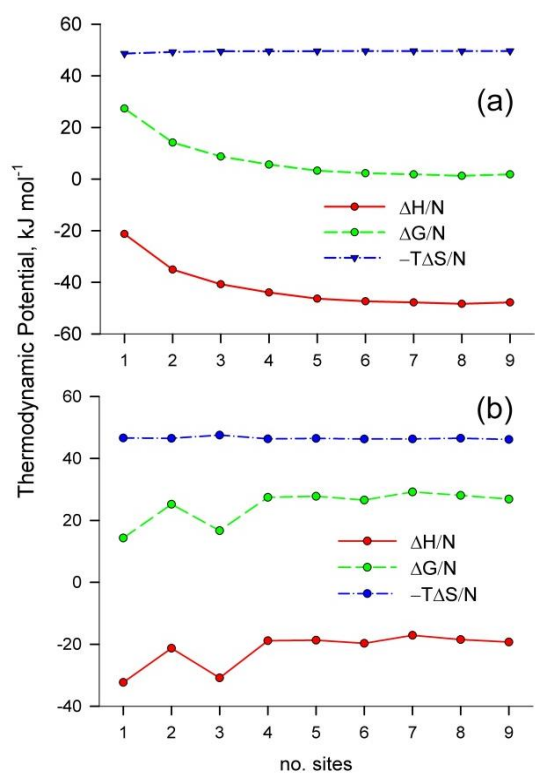

Figure 3. (Color online) Standard thermodynamic potentials of the $C_{60}$ fullerene (a) amino-functionalization and (b) carbamate formation reactions. The depicted potentials were divided by the number of the amino/carboxamido groups grafted.

Fullerene $C_{60}$ exhibits no dipole moment at absolute zero, while thermal fluctuations induce a small dipole moment. In turn, both the amino group and the carboxamido group are polar. Carbamic acid (an intermediate in the production of urea) is more polar than ammonia. Compare 1.4 to 3.2 D for isolated molecules in vacuum. Grafting of these groups to $C_{60}$ introduces substantial dipole moments (Figure 4), which depend on the quantity and location of the polar groups. The largest dipole moment among of amino derivatives, ~5.4 D, was observed in $C_{60}H_4(NH_2)_4$. Interestingly, polycarbamates of $C_{60}$ dot not generally possess higher dipole moments. The dipole moment of $C_{60}H_7(NH_2)_7(CO_2)_7$ is ~6.8 D, whereas the effect of the number of the carboxamido groups is not uniform. Locations of the fullerene sites (Table 1), to which the amino and carboxamido groups are grafted, play a very significant role, even though these groups were distributed more or less uniformly throughout the substrate surface. Seven out of nine simulated polycarboxamidofullerenes exhibit dipole moments between 1 and 3 D. Polarity of the fullerene derivatives is expected to contribute positively to their aqueous solubilities.

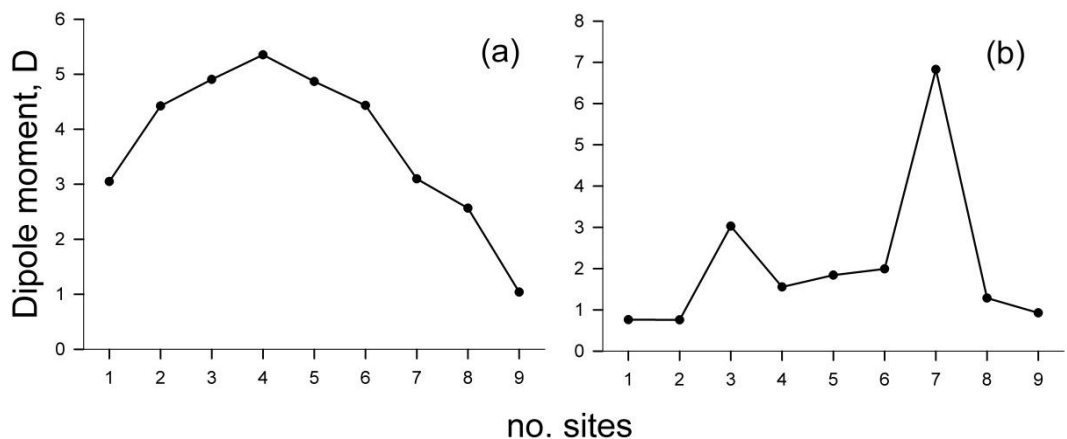

Figure 4. Dipole moment as a function of the number of grafted groups: (a) amino, (b) carboxamido.

Amino- and carboxamido-functionalization of $C_{60}$ changes its HOMO-LUMO band gap substantially (Figure 5). n = 1...4 in $C_{60}H_n(NH_2)_n$ and $C_{60}H_n(NH_2)_n(CO_2)_n$ results in smaller band gaps than in the pristine buckyball, whereas n = 5...9 results in larger band gaps. The band gap increases nearly uniformly as the number of polar groups increases. Step-wise attachment of the amino groups decreases percentage of unsaturated carbon atoms. An effect of this alteration is splitting of the p-orbitals, which are responsible for the HOMO energy. Carbamate formation have a marginal effect on the band gap alteration, while the position of the groups on the fullerene surface (Table 1) matter more significantly. The band gap increase is not strictly uniform. For instance, the band gap in $C_{60}(NH_2)_5$ is slightly higher and $C_{60}(NH_2)_7(CO_2)_7$ is slightly smaller than it could have been expected. While it is difficult to explain small alterations of the band gap in response to each new linked group, we hypothesize that electrostatic interactions between polar grafted groups are responsible for the mentioned effects.

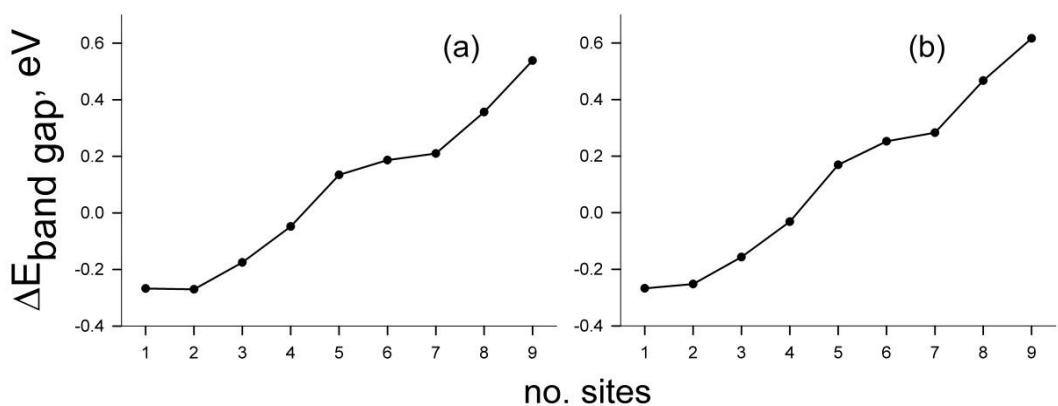

Figure 5. Band gap increase as a function of the number of grafted groups: (a) amino, (b) carboxamido. The experimentally determined band gap of pristine $C_{60}$ amounts to 1.86 eV.[42]

Hydration free energies (Figure 6) become more favorable as the number of polar groups increases. Polyaminofullerenes exhibit a significant affinity to water, whereas the polycarboxamidofullerenes perform even better due to higher polarity of the carbamate moiety. Therefore, an aqueous solubility increases stepwise with capturing each subsequent $CO_2$ molecule. It should be noted that implicit solvation is unable to describe reasonably an entropic factor change upon solvating large molecules. Large absolute errors must be expected, since solvent structure perturbations, which are induced by the solute, are underestimated. The errors are roughly proportional to the size difference between the solute particle and the individual solvent particles. Although the trends revealed by the standard solvation thermodynamics are trustworthy, the reported absolute values must be treated with caution.

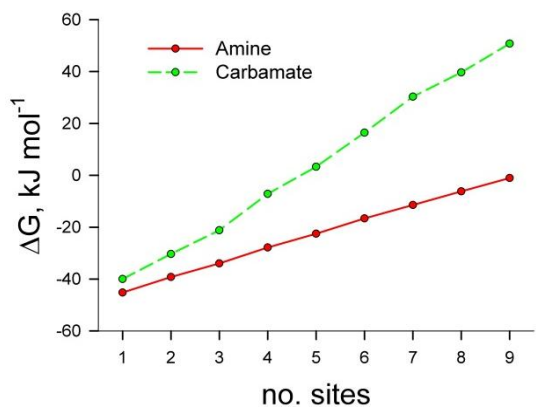

Figure 6. (Color online) Standard free energies of transfer from water to n-octanol of the $C_{60}$ fullerene derivatives computed using an implicit solvent model: amino (red line), carboxamido (green line). Water and n-octanol were used as polar and non-polar molecular solvents, respectively.

The amino group takes (shifts towards itself) a small additional amount of electron density, ~0.1e (Figure 7), by forming a polar covalent bond with $C_{60}$. The amount of the acquired electron density per the amino group only insignificantly depends on the number of these groups. Each carboxamido group takes more electron density than each amino group, since oxygen is more electronegative than nitrogen. In the case of $C_{60}H_9(NH_2)_9(CO_2)_9$, the total Mulliken charge on the backbone is ~1.94e, being 0.24e larger, as compared to $C_{60}H_8(NH_2)_8(CO_2)_8$. Since formation of the carboxamido groups on the fullerene surface leads to a substantial redistribution of electron density, it also adjusts electronic and chemical properties of $C_{60}$. For instance, the polycarboxamido-functionalized fullerene is much more electrophilic and hydrophilic than the pristine $C_{60}$. It must be also understood that an aromatic subsystem of $C_{60}$ is ruined upon polyfunctionalization.

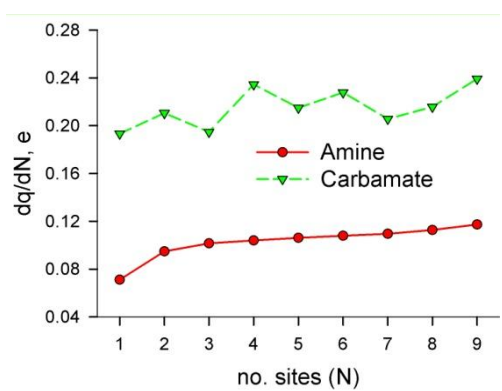

Figure 7. (Color online) Derivative of the total Mulliken charge localized on the fullerene backbone vs. number of the grafted groups: amino (red solid line), carboxamido (green dashed line).

**Conclusions**

Hybrid DFT study was hereby performed to characterize structure, thermodynamic, and electronic properties of the amino- and carboxamido-functionalized fullerene $C_{60}$. The performance of $C_{60}H_n(NH_2)_n$ was found to be essentially independent on the number of $CO_2$ molecules captured. This fact is favorable for using amino-functionalized fullerenes as $CO_2$ scavengers, as soon as the production costs of this or analogous material are lowered enough. While the computed Gibbs free energies for the $CO_2$ fixation reaction are positive, they can be decreased (i.e. made more favorable) by increasing pressure, according to the Le Chatelier's principle.

The dipole moments and band gaps of the amino-functionalized fullerenes are substantially larger than those of pristine $C_{60}$, because p-orbitals are splitted and some double-bonds are destructed upon amino-functionalization. Consequently, such functionalization increases hydrophilicity of the $C_{60}$ fullerene, which may be an important secondary effect. Additionally, chemisorption of $CO_2$ (formation of the carboxamido group) increases hydrophilicity of $C_{60}$. Therefore, as new $CO_2$ molecules are fixed, the functionalized fullerenes become more and more water-soluble. Due to strong electrostatic interactions, the carboxamidofunctionalized C60 may exhibit certain degree of aggregation (cluster formation) in the aqueous solutions. Large-scale classical molecular dynamics simulations may be useful to obtain quantitative insights into such a behavior. A significant fraction of the electron density, up to 1.94e, gets transferred from the $C_{60}$ backbone to the carboxamido groups. The $C_{60}$ backbone becomes positively charged. Polarization of $C_{60}$ is beneficial to adsorb polar gases and promote its coupling with polar environments.

While a lot of cheaper substrates are available to host amino groups, fullerenes constitute definite interest, thanks to their rich chemistry and large surface area. The $C_{60}$ fullerene molecule is compact, only 0.5 nm$^3$, but offers a number of sites for amino-functionalization. If our predictions are confirmed by future experiments, the proposed group of $CO_2$ scavengers may be an interesting alternative to the established $CO_2$ capture schemes. Amino-functionalized carbon nanotubes and graphene may also constitute certain research interest as they offer significant surface area, which can be populated by the amino groups. Investigation of the surface curvature effect on the $CO_2$ chemisorption is an important research avenue.

**Author Information**

E-mail for correspondence: vvchaban@gmail.com (V.V.C.)

**Author Contributions**

N.A.A. conducted quantum-chemical simulations and prepared most figures and tables. V.V.C. formulated research schedule and discussed new results in the context of available literature.